\documentclass[]{emulateapj}

\def\lte{\lower 0.5ex\hbox{${}\buildrel<\over\sim{}$}} 
\def\gte{\lower 0.5ex\hbox{${}\buildrel>\over\sim{}$}} 
\def\lsim{\lower.5ex\hbox{$\; \buildrel < \over \sim\;$}}
\def\gsim{\lower.5ex\hbox{$\; \buildrel > \over \sim\;$}}
\def\psim{\lower.5ex\hbox{$\; \buildrel \propto \over \sim\;$}}
\def\eqb{\begin{equation}}
\def\eqe{\end{equation}}
\def\eqba{\begin{eqnarray}}
\def\eqea{\end{eqnarray}}
\def\bar{\begin{array}}
\def\ear{\end{array}}
\def\no{\nonumber}

\def\Rs{R_{\odot}}
\def\a{\alpha}
\def\b{\beta}

\def\D{\Delta}

\def\th{\theta} 
\def\eps{\varepsilon} 
\def\b{\beta} 
\def\d{\delta} 

\def\Om{\Omega}
\def\Omp0{\Omega_{p_0}}
\def\vB{{\bf B}} 
 
\def\vk{{\bf k}} 

\def\klp{k_{l}}
\def\klp1{k_{l+1}}
\def\kpa{k_{\parallel}}

\def\Vsw{V_{\mbox{{\small sw}}}}
\def\z{\zeta}

\def\tB{\tilde{B}}

\def\kv{{\bf k}}

\def\aref{a_{\mbox{\scriptsize{ref}}}}
\def\Cav{C_{\mbox{{\small av}}}}
\def\Bzav{B_{z_{\mbox{{\tiny av}}}}}
\def\apl{a'_l}
\def\kpl{k_{l}}
\def\kpl1{k_{l+1}}
\def\Cpa{C_{\parallel}}
\def\Lpe{L_{c_{\perp}}}
\def\Lpa{L_{c_{\parallel}}}
\def\ximax{\xi_{\mbox{{\small max}}}}
\def\vBav{{\bf B}_{\mbox{{\small av}}}}

\shorttitle{TURBULENCE-LEVEL VARIABILITY AND ANISOTROPY}
 
\shortauthors{RAGOT} 
 
\begin{document} 
 


\title{TURBULENCE-LEVEL VARIATIONS AND MAGNETIC FIELD ORIENTATIONS 
IN THE FAST SOLAR WIND}
 
\author {B. R. Ragot} 
 
\affil{Helio Research, P.O. Box 1414, Nashua, NH 03061, USA} 
 
 
\begin{abstract} 
The turbulent magnetic fields of a large set of fast solar wind streams 
measured onboard {\em ACE} and {\em STEREO A} and {\em B} are analyzed 
in an effort to identify the effects of the turbulence-level broad variations 
on the orientations of the local, time-averaged magnetic fields. The 
power level of turbulence, roughly defined as the power in the transverse
field fluctuations normalized to the medium-scale average background
field, tightly orders the location of the peaks in the probability 
distribution functions (PDFs) of the angles between local fields and 
Parker spiral. As a result, the broad variations in the power level 
of turbulence cause a steep dependence of the average power level 
of turbulence on the angle of the local field to the Parker spiral, 
with the highest turbulence levels found near the normal to the 
Parker spiral and the lowest levels near the Parker spiral direction. 
Generalized quasilinear estimates of the mean cross-field displacements 
adapted to intermittent time-varying turbulence lead to accurate fits 
of the observed angle PDFs at all stable levels of turbulence, 
supporting the idea that isotropic turbulence 
could account for the observed angle PDFs. Modeling of the 
angles $\a_r$ between local fields and radial direction, from the PDFs of 
the angles $\a$ between local and background fields under an assumption 
of axisymmetry of the turbulent fields around a background field 
in or near the direction of the Parker spiral, also produces fairly 
good fits of the observed PDFs of $\a_r$, thereby validating the assumption. 
Finally, local field reversals are found to be quite common 
even within very broad streams of ``unipolar'' fast solar wind.
\end{abstract}


\section{INTRODUCTION} 

Local spectral analysis of the solar wind (SW) magnetic fields reveals 
strong time variations in the power level of the turbulent field 
fluctuations. This variability of the power level of the field fluctuations 
is responsible for the strong non-Gaussianity of the observed PDFs of 
field variations associated with intermittency (Ragot 2013). 
The strong non-Gaussianity of the observed PDFs of field variations 
is unlikely to be the only observable consequence of the power-level 
variability. Here, we investigate the magnetic field orientations 
as functions of the highly varying power level of turbulence 
in the fast SW, and vice versa, where the power level of turbulence
is defined with the field fluctuations normalized to the medium-scale 
($15\,$min) average background field.

SW magnetic fields $\vB$ are measured along SW flow lines
and the time series of these in situ field measurements 
can be averaged on a timescale $\D t$ to obtain mean 
``magnetic field vectors'' $\vBav(t,\D t)$. If $\d t$ is 
the time resolution of the measurements and $\D t = N \d t$, 
\eqb
\vBav(t, \D t) = {1 \over N+1} \sum_{n=-N/2}^{N/2} \vB(t+n \d t) \; .
\label{Bav}
\eqe
What is the meaning of these mean ``magnetic field vectors''
$\vBav(t, \D t)$? Do they somehow give the direction of 
a real magnetic field within the SW? How do they relate 
to the direction of ``real'' magnetic field lines? 
The answer depends on how much magnetic field lines
diverge from each other, locally, and on how much 
transport they undergo collectively. If the directions
of neighboring field lines strongly deviate from each other 
rather than being dominated by one common direction of 
transport, then it is dubious that the mean ``magnetic
field vectors'' $\vBav$ can be ascribed any other meaning
than that of time averages of in situ magnetic fields along 
SW flow lines. However, if the directions of neighboring 
field lines mostly run ``parallel'' to each other, being
dominated by one common direction of transport, then 
the mean ``magnetic field vectors'' $\vBav$ give the
mean local common direction of transport, or mean local
direction of a field line that passes through one
of the measurement points.

In a magnetic turbulence with background field $\vB_0$, 
such as the SW, both the field-line dispersal (relative
field-line divergence) and global field-line transport 
(common transport) can be assessed theoretically and modeled
(e.g., Ragot 1999, 2006a, 2006b, 2009, 2010a, 2010b, 2011). 
Field-line dispersal and global field-line transport are 
measured by the variations in field-line separations and 
by the field-line displacements across the background-field 
direction, or cross-field displacements, $\D r$, respectively. 
It appears that in slow and fast SW streams, the association 
between time-averaged field vector direction and mean direction 
of an actual field line is justified on most timescales because 
the cross-field displacements $\D r$ far exceed the variations 
in field-line separations on the background field-aligned 
length scales $\D z \le 10^{11}$cm (e.g., Ragot 2010b, 
knowing that $\langle (\D r)^2\rangle^{1/2} \sim \D z$
on these $\D z$ scales).\footnote{A note of caution 
is in order here. Our results concerning the field-line 
dispersal were obtained in self-similar turbulence. The 
stronger fluctuations of the intermittent turbulence at 
the higher frequencies may cause stronger relative field-line 
dispersal. This may need verification in the future. But 
for the time being, we believe that in most cases, the 
variations in field-line separations remain less than 
the cross-field displacements even with the intermittent 
fluctuations of spectral power (see Section 3).}$^,$\footnote{For instance, 
in the slow SW, which is the case for which simulation results have 
been published, the separation between two field lines
that are initially $10^8\,$cm apart may grow tenfold on a 
$\D z$ scale of $10^{11}\,$cm, or even hundredfold on a 
$\D z$ scale of $3\times 10^{11}\,$cm (see Figures 3 and 5 of 
Ragot 2011), and statistically, by factors of $\approx 4.5$ 
and 20 (from Figures 1, 2 and 4 of Ragot 2011), but the variation 
of that separation remains only a small fraction of the cross-field 
displacements undergone by both field lines on these $\D z$ scales, 
here about 0.9 and 3.6 percent at $\D z = 10^{11}$ and 
$3\times 10^{11}\,$cm, respectively, in the flux-tube simulation 
of Ragot (2011), and only 0.4 and 0.7 percent, statistically 
(from Figure~1 of Ragot 2006b, $\langle (\D r)^2\rangle^{1/2} 
\approx 1.1 \times 10^{11}$ and $2.8 \times 10^{11}\,$cm at 
$\D z = 10^{11}$ and $3\times 10^{11}\,$cm, respectively).} 
For a SW speed $\Vsw$, the length scales $\D z$ relate to the 
timescales $\D t$ through $\D z = \Vsw \, \D t \, \cos \phi_r$,
the angle $\phi_r$ between Parker spiral and radial direction 
being defined below in Equation (\ref{spiral}).

The time-averaged field directions obtained by integrating 
in situ fast SW magnetic fields over time intervals of duration 
$\D t \sim 15-20\,$s (depending on $\Vsw$) should therefore match 
the local field-line directions on the background field-aligned 
length scale $\D z \sim 10^9\,$cm. And the angles $\a$ [see 
Equation (\ref{tan_a}) in Section 4] between these time-averaged 
in situ fields and the background field $\vB_0$ should match 
the angles between the mean directions of local individual 
field-line segments of $\vB_0$-projection $\D z$ and $\vB_0$, 
whose tangents are given by the ratios $\D r / \D z$. 
Again, this would not be the case if the field lines were too 
strongly diverging from each other. These premises are at the 
basis of the modeling of Section 4 for the PDFs of the angles 
$\a$ at the scale $\D z = 10^9\,$cm. 

A first-order approximation for the large-scale geometry of the 
SW magnetic fields is given by the Parker-spiral model for 
interplanetary magnetic fields (Parker 1958, 1963). The angle $\phi_r$ 
of the Parker spiral to the radial direction at heliocentric distance 
$r$ is related to the solar rotation rate $\Om_{\Rs}$ and SW speed 
$\Vsw$ through (e.g., Burlaga 1995)
\eqb
\tan \phi_r \approx {r \, \Om_{\Rs} \over \Vsw} \; . 
\label{spiral}
\eqe
The local magnetic fields actually deviate from the Parker spiral 
by large angles, but the Parker spiral direction generally represents
a good axis of symmetry for long statistics of the fields. PDFs of the 
azimuthal angles of the magnetic fields' in-ecliptic projections were 
determined from in situ measurements early on (Ness \& Wilcox 1966; 
Hirshberg 1969; see also, Burlaga \& Ness 1997). More recently, 
PDFs of the angles between SW magnetic fields and Parker spiral 
(or background field) were also determined both in three- and 
two-dimensional space from in situ measurements, and using 
the measured Fourier spectra of magnetic fluctuations, also 
theoretically and numerically (Ragot 2006b). These determinations, 
however, were made from only one single SW stream in each of four 
SW conditions (slow and fast SW at $1$ and $0.3\,$AU), and the time 
variability of the power level of turbulence was not considered. 
Here we focus on the fast SW near $1\,$AU and extend our analysis 
to a large set of streams (over one hundred) observed onboard 
{\em ACE} and {\em STEREO A} and {\em B}. Most importantly, 
we analyze the directions of the local magnetic fields as  
functions of the local level of turbulence, rather than a 
characteristic average value of that level.

An earlier study by Podesta (2009) showed that the turbulent 
energy observed in fast SW streams is on the average higher 
when the local magnetic field is normal to the SW flow. 
The conclusion of that earlier study was that the higher levels 
of turbulent energy seen when the local magnetic field is normal 
to the SW flow are indicative of anisotropic turbulence with a 
wavevector distribution dominated by the wavevectors normal to the field.
We are somewhat skeptical of this conclusion that variations of turbulent 
energy with magnetic field angles necessarily reflect an anisotropy 
of the wavevector distribution. Our skepticism is driven by the facts that
(1) the variations in the turbulent energy and turbulence level are actually 
so much greater than the variations seen in the above anisotropy argument
(orders of magnitude, to be compared to a factor of the order of a few),
and (2) the orientation of the local magnetic field is in large measure
determined by the turbulent energy and turbulence level (see abstract,
next paragraph and Section 3 for a definition of the power level of 
turbulence or turbulence level), as we confirm here through our analysis 
and modeling of the field orientations for a broad range of turbulence levels.

Through our analysis and modeling, we will test whether a turbulence model 
of isotropic wavevector distribution could account for the observed PDFs of 
local field angles at any of the measured turbulence levels. We will also 
test whether the PDFs of angles relative to the radial can be recovered 
from the PDFs of angles relative to the background under an assumption 
of axisymmetry and Parker-spiral background field. 

Here, we distinguish between turbulent energy and turbulence level.
The turbulent energy is basically the power level of the magnetic 
field fluctuations, computed at either one frequency or integrated
over a broader frequency range. By contrast, to anticipate 
our discussion of Section 3.5, the turbulence level is 
the power level of the magnetic field fluctuations {\em normalized 
to the medium-scale average background field.} While the turbulent
energy does say little about the stochasticity of the fields and 
the amount of field-line wandering actually taking place in the SW,
especially if the background field undergoes strong fluctuations,
the turbulence level tightly parametrizes field stochasticity 
and field-line wandering. In fast SW streams, it appears that 
the background field does undergo strong fluctuations. Therefore,
the turbulence level as defined above and in greater detail
in Section 3.5, is a far better ordering parameter for the 
orientations of the fast SW magnetic fields. 
Whether the turbulence level is defined at one frequency, 
integrated over a range of frequencies, or involves a more
complex combination of frequencies through the formula for 
the mean cross-field displacement (see Sections 3.4 and 3.5) 
does not affect that basic fact.

Our data selection is presented in Section 2 and the variations in the 
power level of magnetic fluctuations are shown in Section 3, where the 
power level of intermittent turbulence is also defined. 
In Section 4, through modeling of the PDFs of field-line directions, 
we model the PDFs of the angles $\a$ between local and 
background (or Parker spiral) fields at series of stable levels
of turbulence. Section 5 deals with the transition from the $\a$ angles 
to the angles $\a_r$ between local fields and radial \linebreak direction.
Section 6 presents some consequences of the power variability 
on the field orientations and the \linebreak apparent ``anisotropy'' of 
the turbulence. The conclusion follows in~Section~7.

\section{DATA SELECTION}

The study presented in this paper requires high statistics
of fast SW turbulent magnetic fields. To reach sufficient
statistics, we use data from {\em ACE}, from year 2000 to 2011, 
and from the two {\em STEREO} spacecraft, {\em STEREO Ahead} 
(or {\em A}) and {\em STEREO Behind} (or {\em B}), from year 2007 
to 2012. We selected a total of 121 intervals with SW speeds 
in excess of $500\,$km s$^{-1}$ (but mostly $> 550\,$km s$^{-1}$)
and durations ranging from 1 to 8 days, for a total duration of 307.5 days. 
Stream leading and trailing edges were avoided as far as possible.
Some of the selected intervals obviously belong to the same
SW stream, but for our purpose, they can be considered as 
independent because of the broad separation of the spacecraft
and/or the time lapsed between recurrences. 
$1\,$s-averaged interplanetary magnetic field data from the MAG 
magnetometer on {\em ACE} were downloaded from the ACE Science Center
at www.srl.caltech.edu/ACE/ASC. $125\,$ms-resolution magnetic field data
from the IMPACT/MAG magnetometers on {\em STEREO A} and {\em B} were downloaded 
from www.ssl.berkeley.edu and averaged on 8 consecutive points to match 
the averaging timescale of the {\em ACE} data. In the following, 
all magnetic field data have a time resolution $\d t$ of $1\,$second.

\section{POWER-LEVEL FLUCTUATIONS}

One of the longer intervals selected for our analysis is the 
fast SW stream observed onboard {\em ACE} between day 70 and 76 
of year 2008. We use this interval as an example to illustrate
in Figures 1$-$3 how the Fourier power level of the transverse 
magnetic fields fluctuates with time. 

\subsection{Fourier Spectral Analysis with Sliding Window}

For our analysis, we define the directions $z$, $x$ and $y$ 
as the directions along the background magnetic field $\vB_0$
(or Parker spiral) and normal to $\vB_0$, in the ecliptic plane
($x$) and normal to it ($y$). The angle $\phi_r$ of the Parker 
spiral to the radial direction is determined 
from Equation (\ref{spiral}) using the 
$64\,$s resolution proton speeds from the SWEPAM experiment. 
(For SW streams observed onboard {\em STEREO A} or {\em B}, 
we use the $1\,$min resolution proton bulk speeds from the 
PLASTIC experiment.)\footnote{The analysis of this paper 
has been repeated with proton speeds averaged
over each of the entire SW streams, and over timescales of one day, 
one hour and $15\,$min, with no noticeable variation in the results.
So the exact averaging timescale for Vsw does not seem to affect 
the results of this paper, which probably reflects the low variations 
of $\Vsw$ over each of the fast SW intervals selected for the analysis.} 
We project the magnetic field vectors 
(in RTN coordinates) on these three directions and compute 
the spectra of the $B_x$ and $B_y$ components\footnote{At most 
values of the turbulence parameter $\xi$, formally introduced 
in Section 3.5, the $B_x$ and $B_y$ fluctuations dominate the $B_z$ 
fluctuations and a modeling based on only the $B_x$ and $B_y$ 
fluctuations is appropriate. At the lowest values of $\xi$, where 
$B_z$ fluctuations can become dominant, we will include the effects 
of the``compressive'' field-aligned fluctuations in our modeling 
of the field angles $\alpha$ (see Section 4 and end of 3.4).} 
by fast Fourier transform, with a Hanning window to minimize 
truncation effects. We make this analysis time-dependent by using 
a window of width $T_w \equiv N_w \d t$ and letting the window slide 
by $\d t$ increments. Taking the module of the results, squaring and 
multiplying by the window width $T_w$, we obtain the 
power levels of $B_x$ and $B_y$ before ``smoothing,'' 
\eqba
A_{x,y}(\nu, t, T_w) & \equiv & 
T_w \left|{1 \over N_w} \sum_{n=0}^{N_w-1} {1 \over 2} 
\left[1 - \cos\left({2 \pi n \over N_w}\right)\right] \right. \no \\ 
&  & \hspace*{.6cm} \times \left.
B_{x,y}(t+n \,\d t) \, e^{i \, 2 \pi \, \nu \, n \, \d t / N_w} \right|^2 \!\!\! , 
\label{Axy}
\eqea
and compute their sum, $A \equiv A_x + A_y$. We then reduce the 
noise in the result $A$ by smoothing. For that, we make use 
of the very long-time reference power level, $\aref$, which 
we compute by including the entire interval of data (in this 
particular case, $T_w = 6\,$days), to determine the spectral 
slopes and right the power level $A$ before averaging locally 
in frequencies. The reference power level itself can easily be 
smoothed out at the higher frequencies by averaging over 200 
successive frequencies. In that particular case, the separation 
$\d \nu$ between harmonics is so small that we do not have to worry 
about the spectral slopes when averaging over the neighboring 
frequencies. Because we find that the spectral slopes of $A$ vary 
little with time, we smooth the power level of transverse magnetic 
fluctuations by computing:
\eqb
a(\nu_0, t, T_w) \equiv  {\aref(\nu_0) \over N_s+1} 
\sum_{n_s=-\eps N_s}^{(1-\eps) N_s} 
{A(\nu_0+n_s \d \nu, t, T_w) \over \aref(\nu_0+n_s \d \nu)} \; , 
\label{asmooth}
\eqe
where $0 \le \eps \le 1$ and $\d \nu = 1/T_w$ is the separation 
between Fourier harmonics of $A$, not $\aref$. In Figures 1$-$3, 
$N_s = 200$ and $\eps = 0.15$. 
The smoothing operation is basically a local averaging over the 
frequencies, after correction for the spectral slope using the 
very long-time reference power level, $\aref$. 

With this time-frequency analysis, we are keeping control of both 
the frequency $\nu$ and the window width or time resolution of the 
Fourier analysis, $T_w$. One may argue that a wavelet analysis 
(e.g., Daubechies 1992) is superior, but both Fourier and wavelet 
analysis methods produce similar results (Podesta 2009 used a wavelet 
analysis and our results are consistent, see Section 6). By using a 
Fourier spectral analysis with sliding window, we are just being 
consistent with our earlier work on the wandering of magnetic field lines, 
which is most relevant to the study of the field orientations and whose 
results we are using in Section 4 to model the PDFs of the angles $\a$ 
[see Equation (27) of Ragot 2006a for the definition of the Fourier 
transform used in that earlier work; see also Section 3.4]. In most 
theoretical calculations, it is far easier to Fourier transform 
back and forth than to wavelet analyze. The Fourier analysis 
has for itself its simplicity and ``ease of interpretation.'' 

With this time-frequency analysis, we are not only determining the
power level of the magnetic field fluctuations as a function of time 
(see Sections 3.2 and 3.3), but we are also checking the shapes of 
the frequency spectra. To our analysis of the field directions and 
turbulence it is important that we keep track of the time variations 
or lack thereof of the spectral slopes (see Figure 2 of Ragot 2009 
or Figure 4 of Ragot 2006d). Indeed, the spectral indexes are still 
needed to evaluate the mean field-line directions analytically 
(see Section 3.4). 

\begin{figure}[t]
\epsscale{1.} \plotone{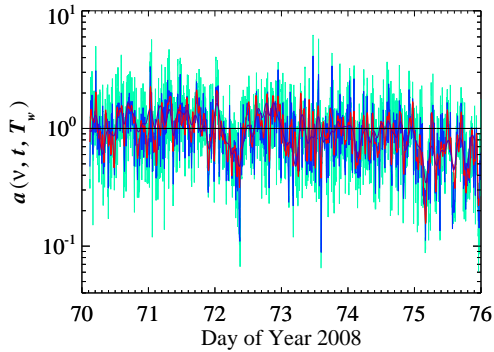} 
\caption{Running power level, $a(\nu, t, T_w)$, at frequency 
$\nu$ and with a time resolution $T_w$, as a function of time, $t$. 
The fast Fourier transforms were computed with a shifting window 
of width $T_w$ equal to the entire SW interval (black line), and $1/30$ 
(violet), $1/100$ (red), $1/300$ (blue), $1/900$ (light blue) of that 
interval duration for a frequency $\nu$ of  $0.122\,$Hz. The running 
power level $a(\nu, t, T_w)$ is smoothed on $200$ successive frequencies 
and normalized in this figure to the very long-time reference power level, $\aref$. } 
\label{fig1} 
\end{figure}

\subsection{Single-Frequency Multiple Time-Resolution Power-Level Fluctuations}

Figure 1 shows the time variations of the power level $a(\nu,t,T_w)$
computed at frequency $\nu \approx 0.122\,$Hz 
with a shifting window of width $T_w$ decreasing from $1/30$ 
the full duration of the SW stream (violet line) down to 
$1/900$ of it (light-blue line) by three successive factors of 
about $1/3$. The black line gives the very long-time reference
power level, $a_{ref}$, to which the running power level $a$ is 
normalized. We see that the shorter the shifting window is, 
that is, the higher the time resolution of the spectral analysis is,
the stronger the fluctuations of the power level are. The higher-resolution
fluctuations far exceed the lower-resolution fluctuations, but roughly
follow them in their time average, fractal-like. 
Figure 2 presents a zoom out of Figure 1, for the first day of data.
The electronic version of Figure 1 can also be zoomed in to follow 
the four fluctuating lines and check how each is fluctuating 
around the line of next lower time resolution.

The power-level fluctuations of Figure 1 are smoothed on 200 successive
frequencies. The effect of that smoothing is to reduce the amplitude
of the fluctuations (see Figures 3 and 4 of Ragot 2013), especially at 
the lower values of power level, which can be most affected by noise. 
We checked that the effect of this reduction in fluctuation amplitude 
is to eliminate the tiny excess ``bump'' at the top of the PDFs of 
field variations presented in Figures 5$-$8 of Ragot (2013). 

\begin{figure}
\epsscale{1.} \plotone{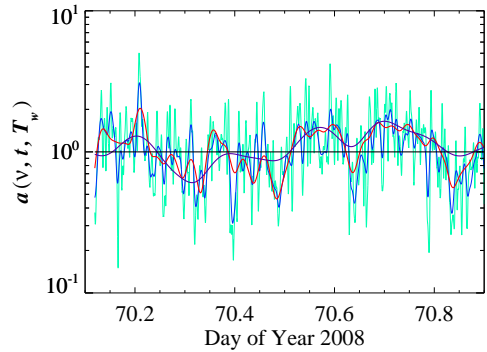} 
\caption{Zoom out of Figure 1, day 70 of year 2008.} 
\label{fig2} 
\end{figure}

\subsection{Multiple-Frequency Power-Level Fluctuations}

\begin{figure}[h]
\epsscale{1.} \plotone{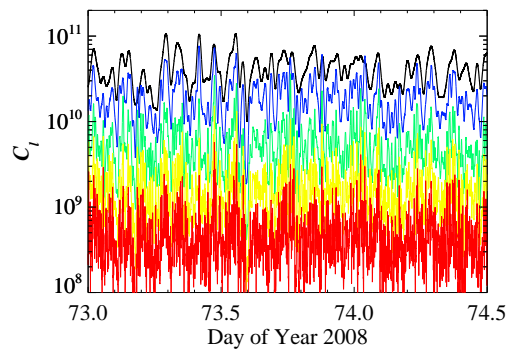} 
\caption{Time variations of the spectral amplitudes 
$C_l \equiv a(\nu_l,t,T_w) \, \Vsw \cos \phi_r$ of transverse
magnetic field, from the lowest wavenumber (black line, $l=0$) to the 
``highest'' (red line, $l=8$) by increments of 2. The amplitudes $C_l$ 
(up to $l=20$) are used to evaluate the turbulence level $\xi$ 
with Equations~(\ref{Dr}), (\ref{Dr_Bz}), (\ref{C}) and~(\ref{xi}).} 
\label{fig3} 
\end{figure} 

When computing the power level at a given frequency $\nu$, the value of 
$\nu$ constrains the width $T_w$ of the window that can be used for the 
Fourier analysis. Typically, $T_w$ has to exceed a few times $\nu^{-1}$,
and in order to capture the time variations of the power, $T_w$ must
also be chosen as short as possible. Because the average slope of the 
turbulence spectrum appears to vary little with time (again, 
see Figure 2 of Ragot 2009 or Figure 4 of Ragot 2006d), and because 
our primary concern here is to capture the precise time variations of 
the power over a broad range of scales, we now compute  
the power at a higher frequency $\nu' \approx 20 \,\nu$ with a window of 
width $T_w = \nu^{-1}$, and extrapolate the result to the lower frequency 
$\nu$ by using the spectral slope obtained with a much longer window, 
covering the entire SW stream. 
Figure 3 shows the resulting time variations of the coefficients 
$C_l = C_0$, $C_2$, $C_4$, $C_6$ and $C_8$ in cm nT$^2$ 
[$a(\nu_l,t,T_w) = C_l / (\Vsw \cos \phi_r)$, see also Section 3.4] 
for the spectral amplitudes of transverse magnetic field, 
starting with the lowest wavenumber 
$k_0 \approx 3.61 \times 10^{-11}\,$cm (black line),
where the turbulence spectrum becomes steeper than $k^{-1}$, 
and ending with $k_8 \approx 7.55 \times 10^{-10}\,$cm (red line).

For a wave vector $\vk$ making an angle $\th$ with the SW flow velocity,
the wavenumber $k$ is related to the frequency $\nu$ through 
$k = 2 \pi \nu / (\Vsw \cos \th)$. Here we are considering 
a series of wave vectors $\vk_l$ along the direction of the background
magnetic field, that is, along the Parker spiral, and corresponding 
to a series of frequencies $\nu_l$. The flow velocity being practically
radial in the SW, the wavenumbers $k_l$ are therefore related to the 
frequencies $\nu_l$ through $k_l = 2 \pi \nu_l / (\Vsw \cos \phi_r)$, 
with $\phi_r$ the angle of the Parker spiral to the radial direction 
at heliocentric distance $r$, as defined by Equation~(\ref{spiral}).
The wavenumbers $k_l$ and wave vectors $\vk_l$ are parallel 
wavenumbers and wave vectors along the Parker spiral, as needed for 
the modeling (see Sections 3.4 and 4), which calls for a one-dimensional 
projected spectrum along the background field [see also Equation (\ref{Cl}) 
and following discussion].


As expected from Figures 1 and 2, in Figure 3 the time variations 
at the higher frequencies (or wavenumbers) roughly follow in their time 
average those of the lower frequencies, but locally display much more 
intense fluctuations. With the time-varying coefficients $C_l$, 
we can estimate in Section 3.4 the magnetic field-line mean 
cross-field displacement, $\langle (\D r)^2 \rangle^{1/2}$, 
in the intermittent turbulence of the SW. 

\subsection{Mean Cross-Field Displacement \\ 
\hspace*{.7cm}in Intermittent Turbulence}

The cross-field displacements $\D r$ on a background field-aligned 
scale $\D z$ are central to our study of magnetic field directions 
because in the turbulence of fast and slow SW streams on the scales
$\D z \le 10^{11}\,$cm, the ratios $\D r/\D z$ are expected to closely 
match (see introduction) the tangents of the angles $\a$ between 
the measured magnetic field, averaged on the corresponding timescale 
$\D t = \D z / (\Vsw \cos \phi_r)$, and the background field $\vB_0$. 
That expectation of a close match between time-averaged field direction, 
obtained from in situ data, and direction of an actual field line is 
justified by our earlier study of field-line dispersal, which 
shows that on the scales $\D z \le 10^{11}$cm in the SW, the cross-field 
displacements far exceed the variations in field-line separations (see, 
e.g., Ragot 2010b, and introduction). Again, this would not be the case 
if the field lines were too strongly diverging from each other.

The cross-field displacement $\D r$ is not defined from the data. 
However, it is a useful quantity whose root mean square, 
$\langle (\D r)^2 \rangle^{1/2}$, 
the mean cross-field displacement, can be evaluated from the 
spectral amplitudes $C_l$ at wavenumber $k_l$ using predictions 
of the generalized quasilinear (GQL) theory (see below). 
$\langle (\D r)^2 \rangle^{1/2}$ is central to our modeling. 
With $\langle (\D r)^2 \rangle^{1/2}$, we can estimate in Section 4
the PDFs of the field-line directions, and compare them to the PDFs 
of the angles $\alpha$ between the time-averaged in situ fields 
and the background field $\vB_0$.

The displacements $\D r$ of turbulent magnetic field lines across 
the direction of a background field $\vB_0$ have been extensively 
studied for a general background field-aligned scale $\D z$
(Ragot 1999, 2006a, 2010a). The square root of their variance 
or mean cross-field displacement, $\langle (\D r)^2 \rangle^{1/2}$, 
can be accurately estimated, both theoretically and numerically,
in self-similar turbulence, that is, in a turbulence of single
``constant-at-all-scales'' spectrum of magnetic fluctuations. 
Both fully nonlinear and GQL estimates can be made 
for $\langle (\D r)^2 \rangle^{1/2}$. On background field-aligned 
scales $\D z \le 10^{11}\,$cm and for turbulence spectral shapes typical
of slow or fast SW streams, nonlinear corrections to the GQL result 
in a background field $\vB_0$
\eqba
\langle (\D r)^2 \rangle^{}_{B_0} & = &  
{4 \over B_0^2} \, \sum_l \left( {C_l \over {(1-\apl) k_l}} \right. \no \\
&  &  \hspace*{-1.9cm} \times 
\left\{1 - |k_l \D z|^{1-\apl} (1-\apl) \, \Gamma(-1+\apl) 
\sin\left({\apl \pi \over 2}\right) \right.  \no \\
&  & \hspace*{-1.4cm} \left. -  F_{P,Q}\left[\left\{{-1+\apl \over 2}\right\}, 
\left\{{1 \over 2}, {1+\apl \over 2}\right\}; 
{-(k_l \D z)^2 \over 4}\right] \right\} \no \\
&  & \hspace*{+1.4cm} - { C_{l+1} \over {(1-\apl) \kpl1}} \no \\
&  &  \hspace*{-1.9cm} \times \left\{
1 - |\kpl1 \D z|^{1-\apl} (1-\apl) \, \Gamma(-1+\apl) 
\sin\left({\apl \pi \over 2}\right) \right. \no \\
&  & \hspace*{-1.4cm} \left. \left.  
-  F_{P,Q}\left[\left\{{-1+\apl \over 2}\right\}, 
\left\{{1 \over 2}, {1+\apl \over 2}\right\}; 
{-(\kpl1 \D z)^2 \over 4}\right] \right\} \right) \no \\
\label{Dr}
\eqea
of Ragot (1999, 2006a) [see also Equation (A13) of 
Ragot~2010a]\footnote{Note that the GQL expression for the mean 
cross-field displacement remains accurate at all scales in the SW 
if $\langle (\D r)^2 \rangle <$ \\[-.05cm]
$2^{1/2} \D z \, \Lpe/\Lpa$ when 
$\D z$ approaches $\Lpa$, $\Lpa$ and $\Lpe$ being the \\[-.05cm]
parallel and perpendicular correlation lengths of the turbulence.}
are found to be negligible, even for much enhanced power levels 
(up to at least a factor of 40 at $10^{11}\,$cm, which would 
easily cover the range of enhancement values observed in Figures 
1$-$3, and much more at shorter $\D z$ scales, see Ragot 2010a). 
On scales $\D z \le 10^{11}\,$cm, accurate estimates 
of the mean cross-field displacement should therefore be obtained
in the intermittent turbulence of the SW with the same GQL expression 
as given by Equation (\ref{Dr}), but with coefficients $C_l$ now 
varying with time and broadly departing from their long-time average 
value, as shown in Figure 3, rather than constant as in a self-similar 
turbulence. The expression of Equation (\ref{Dr}) should only
remain valid, however, as long as strong variations of the 
amplitudes $C_l$ giving the dominant contribution do not occur 
on the timescales shorter than $\D t$, since constant amplitudes
have been assumed on the corresponding lengthscale $\D z$ 
in the integration leading to Equation (\ref{Dr}). We return
to that point in Section 4. 

In Equation (\ref{Dr}), $C_l$ and $a'_l$ are the amplitudes 
$C_l \equiv \Cpa(k_l)$ and spectral indexes of the one-dimensional
or projected Fourier spectrum 
\eqb
\Cpa(\kpa) \equiv \int d\kv_{\perp} \tB_{\perp}^2(\kpa,\kv_{\perp}) 
\label{Cl}
\eqe 
on the background field direction (parallel direction) at parallel 
wavenumbers $\kpa = k_l$, the assumption being that of a piecewise 
power-law spectrum with a well-defined single spectral index 
on each interval $[k_l, k_{l+1}]$. In Equation (\ref{Cl}),
$\tB_{\perp}$ denotes the theoretical three-dimensional Fourier 
transform (with sliding window) of the transverse component 
of magnetic field. 

Equation (\ref{Dr}) 
applies whatever the three-dimensional distribution of wavevectors, 
as long as it produces the projected spectrum of Equation (\ref{Cl}). 
The SW measurements, however, give the spectrum projected 
on the SW flow direction rather than the background field. So the 
measured amplitudes $C_l$ of Figure 3 may not strictly coincide with the 
required $\Cpa(k_l)$, and the spectral indices $a'_l$ used in our further 
estimates of $\langle (\D r)^2 \rangle$ are not necessarily the spectral 
indexes of the projected Fourier spectrum of Equation (\ref{Cl}). Only if 
the three-dimensional distribution of wavevectors is nearly isotropic 
(as found in slow SW by Narita et al. 2010 at the scale of $10^9\,$cm) 
should both sets of amplitudes and spectral indexes \linebreak 
a priori coincide.
The anisotropy found in the Cluster data analysis of Narita et al. (2010) 
becomes significant below the scale of $10^9\,$cm, with a $\sim \pm 50$ 
percent variation in the energy distribution from the parallel to the 
perpendicular directions at $4 \times 10^8\,$cm, but these shorter 
turbulent scales do not contribute (or only very little) to the field-line 
displacements at the scales $\D z \sim 10^9\,$cm considered in this paper, 
nor to the mean cross-field displacements at any other scale. 
We therefore are hopeful that the discrepancy in the measurement 
direction of the one-dimensional spectra will not become an issue 
as we further make use of Equation (\ref{Dr}) in our analysis of \linebreak 
Section 4.\footnote{Put in other words, if the turbulence were 
strongly anisotropic, we should not be able to fit in Section 4 
the PDFs of the field directions using the coefficients $C_l$ 
and spectral indices $a'_l$ obtained from measurements made 
along a direction that differs from that of the Parker spiral. 
But we do obtain quite reasonable fits (see Figures 8, 10 and 11
in Section 4).}

In Equation (\ref{Dr}), $F_{P,Q}$ denotes the hypergeometric function,
computed as a generalized hypergeometric series 
\eqba
F_{P,Q}[\{\a\}, \{\b, \eta\}; \z] & \equiv & 1 + {\a \, \z \over \b \, \eta} + 
{\a (\a+1) \, \z^2 \over \b (\b+1) \, \eta (\eta+1) \, 2!} \no \\
 &  & \hspace*{-2.3cm} + {\a (\a+1) (\a+2) \, \z^3 \over 
\b (\b+1) (\b+2) \, \eta (\eta+1) (\eta+2) \, 3!} + \dots
\label{FPQ}
\eqea
(e.g., Gradshteyn \& Ryzhik 1980) and $\Gamma$ is the Gamma function 
(Euler's integral of the second kind) defined for $\mbox{Re}\, \a > 0$
by the integral 
\eqb
\Gamma(\a) \equiv \int_0^{+\infty} dx \, e^{-x} x^{\a-1} \; . 
\label{Gamma}
\eqe 

The GQL result of Equation (\ref{Dr}) assumes a constant background field 
of magnitude $|B_0|$. However, the starting equation for the field-line
tangent (see Equation (1) of Ragot 1999 or 2006a) calls for the actual $B_z$ 
field projection on the background $\vB_0$, and provided that $B_z$ does not 
fluctuate too fast, or is averaged on a timescale long enough, it can 
obviously substitute for $B_0$ in Equation~(\ref{Dr}). When there are 
significant long-scale fluctuations in the value of $B_z$, that substitution 
should greatly improve the estimate of the mean cross-field displacement. 
Therefore we write the mean cross-field displacement as 
\eqb
\langle (\D r)^2 \rangle^{1/2} = \langle (\D r)^2 \rangle^{1/2}_{\Bzav} 
\equiv {|B_0| \over |\Bzav|} \, \langle (\D r)^2 \rangle^{1/2}_{B_0} \; ,
\label{Dr_Bz}
\eqe
where $\langle (\D r)^2 \rangle^{}_{B_0}$ is given by Equation (\ref{Dr}).
Throughout the rest of this paper, unless otherwise specified, 
$\langle (\D r)^2 \rangle^{1/2}$ denotes $\langle (\D r)^2 \rangle^{1/2}_{\Bzav}$,
the subscript $\Bzav$ of the medium-scale averaged $B_z$-field being 
omitted to lighten the notations. The background field 
$B_0$, or very long-time stream-average field, is computed for each of 
our fast SW streams by averaging the $B_z$-field over the entire stream. 
Shorter-scale fluctuations $\d B_z$ of $B_z$ can be taken into account 
perturbatively in the evaluation of the mean cross-field displacement
by expanding $[1+(\d B_z/B_0)^2]^{-1}$, which we later do to estimate 
the correction due to ``compressibility effects.'' An appropriate 
medium scale for averaging the background field and computing $\Bzav$
remains to be chosen. We return to the averaging scale of the 
local background field $\Bzav$ in Section~3.6. 

\subsection{Defining a Turbulence Level 
for the Time-Varying Intermittent Turbulence}

From the multiple time-resolution and multiple-frequency Fourier 
spectral analysis of Sections 3.2 and 3.3, we realize that 
the ``power level'' of the intermittent SW turbulence is an 
extremely complex function of time and frequency, with wild 
fluctuations of large amplitude. In our study of the magnetic 
field orientations with varying turbulence level, we are 
therefore faced with a first serious challenge that consists 
in defining a turbulence level. 

Our choice of a proper physical parameter that would both be 
a good representation of the level of turbulence at a given time 
and provide a good parametrization of the field orientations 
is guided here by our earlier theoretical and numerical studies 
of magnetic field-line wandering (Ragot 1999, 2006a, 2010). We know from 
these earlier studies that the displacements $\D r$ across the direction 
of the background field $\vB_0$ on a given field-aligned scale $\D z$, 
which should give good estimates of the directions of the in situ 
magnetic fields averaged on the corresponding timescale 
$\D t = \D z / (\Vsw \cos \phi_r)$, is not dependent on the Fourier 
power level at just one frequency, but at a broad range of frequencies 
[see Section 3.4 and Equation (\ref{Dr})]. Because the spectral power level 
is so wildly varying with time and frequency (see Figures 1--3), studying 
the field directions as functions of the spectral power level at any one 
single frequency may therefore not reveal the most meaningful trends of 
the data. By instead studying the field directions as functions of 
a linear combination of the Fourier power levels $C_l$ at a broad 
range of frequencies (or wavenumbers) that happen to produce the 
variance $\langle (\D r)^2 \rangle$ of the field-line displacements 
[see Equations (\ref{Dr}) and (\ref{Dr_Bz})], we believe that we are 
optimizing our chances of uncovering the underlying organization of 
the field directions, and of modeling these field directions successfully. 

The use of the mean cross-field displacement, 
$\langle (\D r)^2 \rangle^{1/2}$, which is a linear combination of 
the spectral amplitudes $C_l$, rather than that of just one of the 
spectral amplitudes $C_l$, presents the advantage of incorporating 
the variability effects at all frequencies. Also, unlike the spectral 
amplitudes $C_l$, which may vary by up to a couple of orders of magnitude 
with time but actually tell us little about the turbulent behavior of 
the fields, the mean cross-field displacement has an obvious simple physical 
meaning that is directly related to the turbulent behavior of the fields. 
It is the quantity that we will be using here to quantify the turbulence 
level. More precisely, we introduce the parameter $C$ defined by 
\eqb
C^{1/2} \equiv 
{\langle (\D r)^2 \rangle^{1/2}_{\Bzav} \over \D z}\; ,
\label{C}
\eqe
with the mean cross-field displacement $\langle (\D r)^2 \rangle^{1/2}_{\Bzav}$ 
given by Equations (\ref{Dr_Bz}) and (\ref{Dr}). 
The quantity $C^{1/2}$ is the mean cross-field displacement
estimated with a background field magnitude equal to the
average $|\Bzav|$ and divided by the field-aligned scale $\D z$.
The parameter $C$ is a linear combination of the spectral amplitudes 
$C_l$ of magnetic field fluctuations [see Equation (\ref{Dr})]. 
The product $C \, \D z$ is not a diffusion coefficient because 
the fields do not diffuse on the scales $\D z \le 10^{11}\,$cm.
Also, due to intermittency, $C$ strongly varies with time.
A related parameter, which we will be abundantly using 
throughout this paper, is the angle 
\eqb
\xi \equiv \mbox{arctan}\left(C^{1/2}\right) \; .
\label{xi}
\eqe
Note that if the turbulence were self-similar with only one 
``constant-at-all-scales'' spectrum of magnetic fluctuations, 
then $\langle (\D r)^2 \rangle$ and $C$ would simply be 
proportional to any one of the amplitudes $C_l$, that is, 
to what is ordinarily understood as the energy level of 
turbulence in self-similar, non-intermittent turbulence. 
It is only because the SW turbulence is strongly intermittent 
that involving the mean cross-field displacement 
$\langle (\D r)^2 \rangle^{1/2}$ in the definition of
the turbulence level presents an advantage.

\begin{figure} 
\epsscale{1.} \plotone{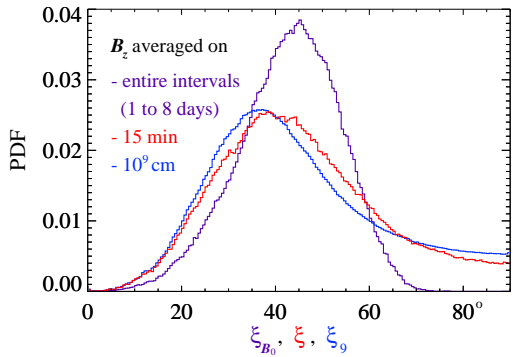} \vspace*{-.25cm}
\caption{PDFs of $\xi_{_{B_0}} \equiv 
\mbox{arctan}[\langle(\D r)^2\rangle^{1/2}_{B_0}/\D z]$ 
(violet), $\xi \equiv $ 
$\mbox{arctan}[\langle (\D r)^2 \rangle^{1/2}_{\Bzav} / \D z]$ (red) and 
$\xi_{_{\huge 9}} \equiv \mbox{arctan}[(B_0 / B_{z_{10^9\mbox{{\tiny cm}}}})$
$\langle (\D r)^2 \rangle^{1/2}_{B_0} / \D z]$ (blue) for entire data set 
of fast SW streams. The long-time average field $B_0$ is computed for 
each individual SW stream (1 to 8 days in duration) while the average field 
$\Bzav$ is the running average over a timescale of $15\,$min. \\[-.1cm]} 
\label{fig4} 
\end{figure} 

\subsection{Turbulence Level, $B_z$-Field Averaging Scale \\
and Field Reversals}

An appropriate scale for computing the local average background field 
$\Bzav$ remains to be chosen. The PDFs of the turbulence-level 
parameter $\xi$, computed for a $B_z$ field averaged over each 
complete SW stream (violet), and over scales of $15\,$min (red) 
and $10^9\,$cm (blue), are presented in Figure 4 for our entire 
data set (121 intervals, 307.5 days).\footnote{For each averaging
scale, we average the $B_z$-component of the field, not its 
magnitude. When averaged on a complete SW stream to obtain $B_0$, 
the $B_z$-component does not ``cancel out'' because we are 
considering fast SW streams of mostly unipolar fields.} [The values of 
$\xi$ are computed from the $C_l$ amplitudes of Figure 3 using Equations
(\ref{Dr}) and (\ref{Dr_Bz}).] A complete SW stream is much too long 
(one to 8 days), while $10^9\,$cm, the scale over which we are computing 
the mean cross-field displacement, is too short. We find that
an averaging scale of $15\,$min is a good compromise between 
the two. The PDFs of $\xi \equiv \xi_{15\,\mbox{{\small min}}}$ and 
$\xi_9 \equiv  \xi_{10^9\,\mbox{{\small cm}}}$ differ relatively
little from each other, meaning that a $15\,$min average already
captures most background field fluctuations. Also, a scale
of $15\,$min in fast SW of speed exceeding $500-600\,$km s$^{-1}$
is a safe 40 to 50 times longer than the scale $\D z = 10^9\,$cm.
In the  rest of this paper, unless otherwise specified,
$\Bzav$ denotes the $15\,$min average of $B_z$. 

\begin{figure} 
\epsscale{1.08} \plotone{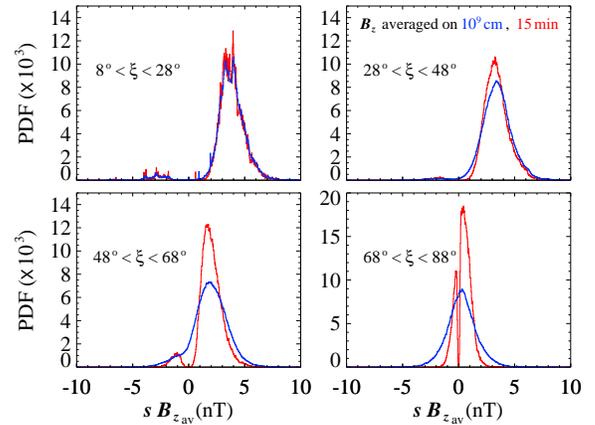} \vspace*{.1cm}
\caption{PDFs of $s \Bzav$ for four intervals of turbulence levels  
($8\degr < \xi < 28\degr$, $28\degr < \xi < 48\degr$, 
$48\degr < \xi < 68\degr$ and $68\degr < \xi < 88\degr$) and a
$B_z$-field averaged on $10^9\,$cm (blue) and $15\,$min (red).
$s$~is the sign of the background field polarization. The lower panels 
show an increasing rate of short-scale field reversals at increasing 
turbulence levels. \\[-.1cm]}
\label{fig5} 
\end{figure} 

Figure 5 now compares the PDFs of $s \Bzav$, where $\Bzav$ is averaged
on scales of $10^9\,$cm (blue) and $15\,$min (red), in four
contiguous intervals of the parameter $\xi$ covering a total
of 80\degr. The parameter $\xi$ is estimated with an averaging 
scale of $15\,$min for $\Bzav$ for both the blue and red PDFs. 
The polarization sign $s$ of the background field is computed 
from each complete SW stream. A negative value for $s \Bzav$ indicates 
a reversal of the field relative to the unperturbed Parker spiral 
field. While reversals are rare events at $\xi < 48$\degr, they 
become relatively frequent at the higher values of $\xi$. At the 
highest levels of turbulence ($\xi > 68$\degr, bottom-right panel of 
Figure 5), short-scale reversals are so frequent that on the shorter 
scale of $10^9\,$cm, reversed fields are practically as common as 
normal-polarity fields. This of course does not mean that field 
reversals are ubiquitous throughout the fast SW. The occurrence rate 
of $\xi > 68\degr$ is relatively low (see red curve in Figure 4). Still, 
it is far from negligible, and we find that if the unperturbed field is 
indeed along the Parker spiral direction, then field reversals are not 
rare occurrences in the fast SW. On the whole, again if the unperturbed 
field is along the Parker spiral direction, the field appears to be 
reversed between 4 and 10 percent of the time in fast SW streams. 
Note also that unlike for the $10^9$cm-average, the PDFs 
for the $15\,$min-average fall to zero at $s \Bzav=0$. 

\section{MODELING OF THE LOCAL-TO-BACKGROUND \\ FIELD ANGLES}

From the mean cross-field displacement of Equations (\ref{Dr}) and 
(\ref{Dr_Bz}), we can model the PDF of the angles $\a$ between local 
and background fields.\footnote{In all rigor, 
we should be using different notations for the angles $\a$
obtained from the model and from the measurements, since they 
could be very different quantities, were it not for the fact that
in SW streams and on the scales $\D z \le 10^{11}\,$cm, 
the cross-field displacements far exceed the variations in 
field-line separations (see intoduction and beginning of Section 3.4).
The angles of the model are the angles $\a_{FL}$ between segments
of magnetic field lines with projection $\D z$ on the background field,
and the background field, while the measured angles are the angles $\a$
between the time-averaged in situ fields and the background field $\vB_0$
or Parker spiral. These facts being clear, we will be omitting the 
subscript $FL$ of the angles $\a_{FL}$. } 
As long as the GQL \linebreak result holds, we expect Gaussian distributions 
for the displacements $\D x$ and $\D y$ in each of the directions 
perpendicular to the background field and a distribution, 
\eqb
f_R(\D r) = {2 \D r \over \langle (\D r)^2 \rangle} 
e^{-(\D r)^2/\langle (\D r)^2 \rangle} \; ,
\label{fR}
\eqe
for the cross-field displacements 
$\D r \! = \! [(\D x)^2+(\D y)^2]^{1/2}$. From this distribution, 
we derive in Section 4.2 of Ragot (2006b) the PDF,
\eqb
f_{A}(\a) =  
{\D z \over \cos^2 \a} f_R\left[(\D z) \tan \a \right] \; ,
\label{fTh}
\eqe
for the angles $\a$ between local and background fields. 

\begin{figure} 
\epsscale{1.} \plotone{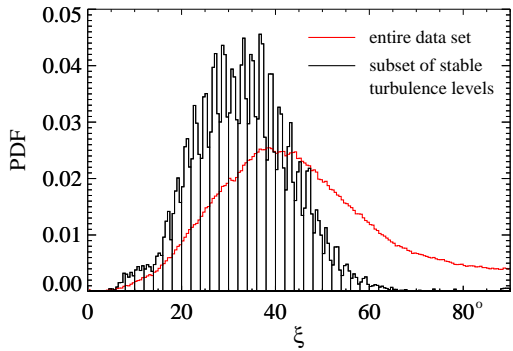} \vspace*{-.3cm}
\caption{PDFs of $\xi \equiv \mbox{arctan}[\langle (\D r)^2 \rangle^{1/2}_{\Bzav}
 / \D z]$ for entire data set (red) and for a subset of time intervals
over which the turbulence level, characterized by $\xi$, remains stable
on a timescale of at least $400\,$s, that is, remains within the same 
2\degr-wide bin over that timescale. The contributions from the highest 
turbulence levels are much reduced in the ``stable'' subset because 
the peaks of higher turbulence levels are usually of very short duration.} 
\label{fig6} 
\end{figure} 

\vspace*{.4cm}
The GQL expression for the mean cross-field displacement
in Equation (\ref{Dr}) assumes a turbulence spectrum 
that does not vary on the scale $\D z$ over which 
$\langle (\D r)^2 \rangle$ is estimated. For a fixed 
value of the parameter $\langle (\D r)^2 \rangle$, 
one should therefore
expect our theoretical prediction to fit the observations
only when the power level is stable on the scale $\D z$. 
From our large data set, we extract all the time intervals
over which such conditions are satisfied. In Figure 6, 
we show the PDFs of the ``power level'' $\xi$ for both
the entire data set (red) and the subset of time intervals
over which the ``power level'' is stable, that is, the subset 
of time intervals over which $\xi$ remains within one 2\degr-wide
bin for a time exceeding $400\,$s (which is slightly larger than 
$20 \, \D z / \Vsw$). The contribution from the 
highest turbulence levels are much reduced in Figure 6
in the ``stable'' subset because the peaks of higher power
are usually of short duration or involve very steep time variations. 

\begin{figure} 
\epsscale{1.} \plotone{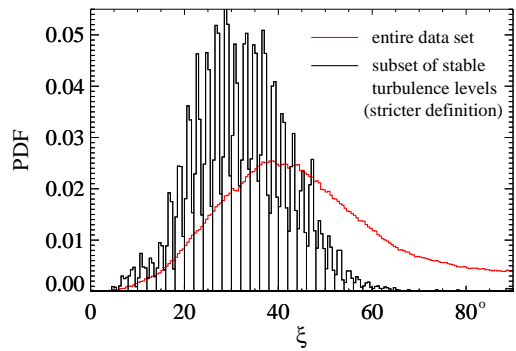} \vspace*{-.2cm}
\caption{Same as Figure 6, but for a stricter definition of the intervals
of ``stability,'' wherein $150\,$s are removed at the ends of each 
subinterval. The contribution of each subinterval is now more peaked 
around its center.} 
\label{fig7} 
\end{figure} 

\vspace*{.1cm}
Figure 7 shows similar PDFs for a stricter definition of stability 
whereby the edges of the stability intervals are removed (some 
$150\,$s on each end). This eliminates time intervals over which 
the computed turbulence power level may still be influenced by time 
intervals with different dominant power level. The contribution of each 
subinterval is now more peaked around its center, but the PDF shape 
is not otherwise significantly modified. 

The purpose of these subsets of stable turbulence power levels
is to check whether the angles between \linebreak local and background fields
can be accurately modeled and understood when the basic model requirement
of stable turbulence levels is satisfied. For more generally varying 
turbulence power levels, the modeling would require a more complex 
convolution to include the contributions from all power levels. This is 
beyond the scope of the current paper. Here, we just want to find out 
whether the model has a chance to work in the first place and check the 
basics. We thus deconvolve the problem by decomposing the data into 
distinct turbulence power levels, instead of convolving the model. 

On the $\D z$ scales less than $10^{11}$cm, the cross-field displacements 
far exceed the variations in field-line separations (see, e.g., Ragot 2010b).
The cross-field displacements obtained by integrating the $B_x$ and $B_y$ 
measured in situ field components over time intervals of duration 
$\D t$ should therefore match the cross-field displacements of real magnetic 
field lines relatively well, and give good estimates of the actual angles 
$\a$ between local and background magnetic fields in the SW. Again, 
this would not be the case if the field lines were too strongly diverging 
from each other (see introduction, including footnotes). 

\begin{figure} 
\epsscale{1.15} \hspace*{-.1cm} \plotone{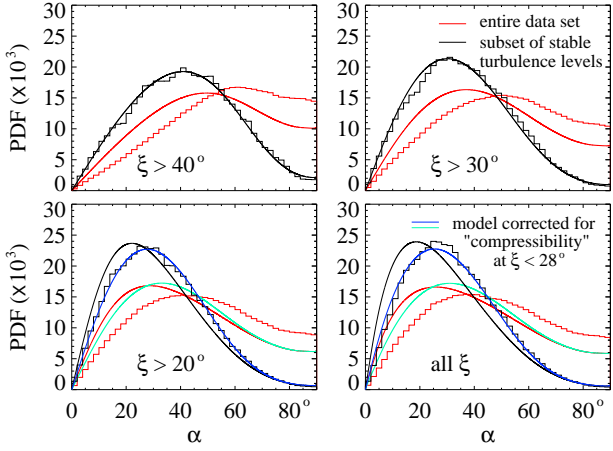} \vspace*{.15cm}
\caption{Measured PDFs (histograms) and model PDFs (lines) of 
the angle $\a$ between local and background fields for entire 
data set (red) and for subset of stable turbulence levels (black) 
for turbulence levels $\xi > 40\degr$ (top-left panel), $\xi > 30\degr$ 
(top-right panel), $\xi > 20\degr$ (bottom-left panel) and all turbulence
levels (bottom-right panel). The blue lines in the bottom panels are 
the model PDFs corrected for the measured ``compressibility'' of 
the turbulence. The model produces good fits for the subset of
stable turbulence levels, but as expected, fails to fit the PDFs
for the entire data set that includes intervals of fast varying 
turbulence level.} 
\label{fig8} 
\end{figure} 
 
In Figure 8, we compare the model to the measured PDFs of the angle
$\a$ between local and background magnetic fields (see footnote 8).
The measured angles $\a$ are obtained from  
\eqb
\tan \a =  
{\left\{\!\left[\sum_{-N/2}^{N/2} \! B_x(t\!+\!n\,\d t)\right]^2 \!\!\!\! + \!
\left[\sum_{-N/2}^{N/2} \! B_y(t\!+\!n\,\d t)\right]^2\!\right\}^{\!1/2} \over 
s \; \sum_{n=-N/2}^{N/2} B_z(t+n\,\d t)} \, , 
\label{tan_a}
\eqe
with the background field polarity sign, $s \equiv B_0/|B_0|$, for each 
sampling time $t$, excluding the first and last $N/2 = \D t / (2 \d t)$ 
measurements of each SW stream (see also footnote 7). 
The value of $N$ slightly varies with time. 
It is given by the ratio of $\D z = 10^9\,$cm over the local product 
$\Vsw \, \d t \, \cos \phi_r$ at time $t$. [If $N$ is an odd number, 
the sums are taken from $-(N-1)/2$ to $(N+1)/2$.] The PDFs computed 
from the entire data set are shown in red, while those computed for the 
subset of intervals with stable turbulence levels are shown in black. 
In each of the four panels, a different range of turbulence
levels is included, from $\xi > 40\degr$ in the upper-left panel
to all $\xi$ values in the lower-right panel. The reason for this
distinction is that at the lowest values of $\xi$, typically below 
28\degr, ``compressive'' field-aligned fluctuations become dominant. 
To the ``non-compressive'' transverse model based on Equation (\ref{Dr}), 
we then have to add a correction. This is done by expanding 
$[1+(\d B_z/B_0)^2]^{-1}$, and roughly results in the multiplication
of the transverse result for $\langle (\D r)^2 \rangle$
by a factor $1+3 \langle (\d B_z/B_0)^2 \rangle$. Numerically,
it amounts to a factor of the order of $1.5$ in front of $\Bzav$.
The results for the PDFs of $\a$ are the blue lines shown 
in the bottom panels of Figure 8. 

The model (black lines in upper panels and dark-blue lines in bottom panels) 
fits the measured PDFs well for the subset of stable turbulence levels,
confirming the validity of the model. As expected, however, the model does 
not fit the measured PDFs so accurately (see red and light-blue PDFs) when 
intervals of fast varying turbulence level are included in the data.

We now decompose the $\a$ PDFs into four ranges of turbulence levels
($8\degr < \xi < 28\degr$ in violet, $28\degr < \xi < 48\degr$ in green, 
$48\degr < \xi < 68\degr$ in blue and $68\degr < \xi < 88\degr$ in red). 
In Figure 9, we compare the PDFs of each subrange for the subset of 
stable turbulence levels and for the entire data set, both for a 
$B_z$-field averaged on $10^9\,$cm (top panel) and on $15\,$min [bottom panel, 
where the sum in the denominator of Equation (\ref{tan_a}) is made from 
-450 to 450 instead of -N/2 to N/2] in the definition of $\tan \a$. 
The measured $\a$ PDFs differ little at the lower turbulence levels, 
but more significantly at the higher levels.

\begin{figure} 
\epsscale{1.1} \hspace*{.2cm} \plotone{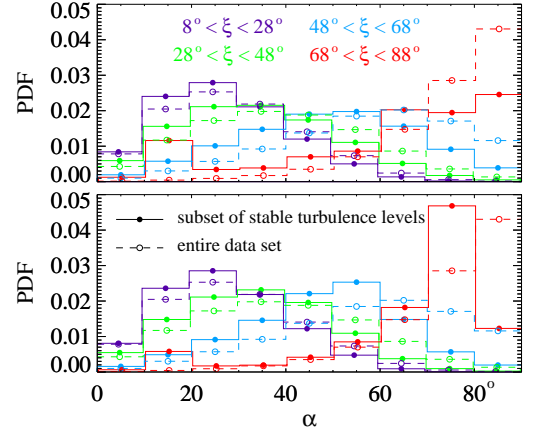} \vspace*{.2cm}
\caption{Measured PDFs of $\a$ for entire data set (dashed-line histograms
with empty circles) and for subset of stable turbulence levels 
(solid-line histograms with filled circles) decomposed into subintervals
of turbulence levels, $8\degr < \xi < 28\degr$ (violet), 
$28\degr < \xi < 48\degr$ (green), $48\degr < \xi < 68\degr$ (blue), 
$68\degr < \xi < 88\degr$ (red), for $10^9\,$cm (top) and 
$15\,$min (bottom) averaged $B_z$ field in the definition of $\tan \a$. \\[-.3cm]} 
\label{fig9} 
\end{figure} 

\begin{figure} 
\epsscale{1.} \plotone{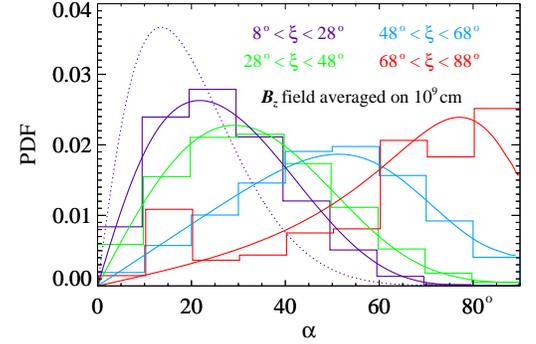} \vspace*{-.3cm}
\caption{Measured PDFs (histograms) and model PDFs (lines) of 
the angle $\a$ between local and background fields for subset of 
stable turbulence levels decomposed into four subintervals
of turbulence levels, $8\degr < \xi < 28\degr$ (violet), 
$28\degr < \xi < 48\degr$ (green), $48\degr < \xi < 68\degr$ (blue), 
$68\degr < \xi < 88\degr$ (red), for $10^9\,$cm-averaged $B_z$ field.
The violet dotted line shows the model PDF at low turbulence level 
prior to correction for ``compressibility effect.'' \\[-.3cm]} 
\label{fig10} 
\end{figure} 

\begin{figure} 
\epsscale{1.} \plotone{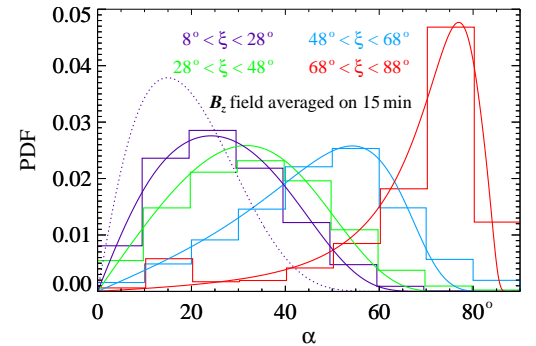}  \vspace*{-.3cm}
\caption{Same as Fig. 10, but for a $15\,$min-averaged $B_z$ field.} 
\label{fig11} 
\end{figure} 

With the same decomposition into four ranges of turbulence levels, 
we compare in Figure 10 the measured normalized $\a$ PDFs of the 
stable subset with the model PDFs. The statistics in the highest
$\xi$-range are very limited and the error bars on the measured PDF 
very large.\footnote{It is the main reason for the high statistics 
of our data set.} We therefore consider the model a reasonable fit
in all four ranges of turbulence levels. While in Figure 10, 
the angles $\a$ are computed with a $B_z$-field averaged 
on $10^9\,$cm, in Figure 11 the $B_z$-field is averaged on a scale
of $15\,$min in the denominator of Equation (\ref{tan_a})
(sum made from -450 to 450 instead of -N/2 to N/2). 
The comparison data/modeling remains satisfactory.

Figures 8, 10 and 11 reveal no problem with our modeling of the 
magnetic fields and of their orientations relative to the background
field, as long as the turbulence level varies sufficiently slowly. 
On all subsets of stable turbulence level, we obtain 
fits of the measured PDFs that are actually quite good, even at the 
highest turbulence levels where the theory was not originally intended. 
Our GQL modeling and results do not assume anything about the turbulence 
isotropy or lack thereof. The only requirement for our turbulence model 
is that the three-dimensional distribution of wavevectors produces the 
right projected spectrum of turbulence [see Equation (\ref{Cl})]. So 
the positive result of our modeling actually supports an isotropic model 
of turbulence by not excluding it.

\section{FROM $\a$ TO $\a_r$}

Because the direction of the radial is known while that of the
background field is in some measure assumed, the orientations
of the SW magnetic fields are more often studied relative to 
the radial than to the background. Here, we go from the 
angles $\a$ relative to the background to the angles $\a_r$
relative to the radial, compare the PDFs of the two at a number
of turbulence levels, and investigate how one goes from 
one to the other. 

Returning to the entire data set, we show in Figure~12 
the PDFs of field angles at a series of turbulence levels
$\xi = 11 \pm 1$\degr, $\xi = 31 \pm 1$\degr, $\xi = 51 \pm 1$\degr, 
$\xi = 71 \pm 1$\degr\ and $\xi = 89 \pm 1$\degr. In Figure~12, 
only the central PDF at $\xi = 51 \pm 1$\degr\ is normalized.
The other PDFs are given relative to that PDF and have much 
lower integrals at the lower and higher ends of the turbulence
power levels. The PDFs of measured $\a$ angles between 
local and background fields are shown in dashed lines while
the PDFs of measured $\a_r$ angles between local field 
and radial direction are shown in solid lines. 
The measured angles $\a_r$ are obtained from
\eqb
\tan \a_r =  
{\left\{\!\left[\sum_{\mbox{-}N/2}^{N/2} \! B_T(t\!+\!n\,\d t)\right]^2 
\!\!\!\! + \! \left[\sum_{\mbox{-}N/2}^{N/2} \! 
B_N(t\!+\!n\,\d t)\right]^2\!\right\}^{\!1/2} \over 
({B_{0_R} / |B_{0_R}|}) \, \sum_{n=\mbox{-}N/2}^{N/2} B_R(t+n\,\d t)} \, , 
\label{tan_ar}
\eqe
where $B_R$, $B_T$ and $B_N$ are the measured magnetic field 
components in the RTN coordinate system ($R$: along the radial,
pointing away from the Sun; $T$: orthoradial in the ecliptic, 
pointing west; $N$: normal to the ecliptic, pointing north). 
$B_{0_R}$ is the R-component of the very long-time, stream-average 
field, the ratio $B_{0_R} / |B_{0_R}|$ giving again the field 
polarity in each SW stream.

To test our understanding of the measured fields and a couple
of simple hypotheses, we now use the PDFs of measured $\a$ 
to simulate the angles $\a_r$ and their PDFs. Through Monte-Carlo  
simulation, we generate sets of $\a$ angles with the measured PDFs. 
Assuming (1) a background field along the Parker spiral direction 
and (2) axisymmetry of the turbulent fields around that background
field, we then transform the angles $\a$ into the angles $\a_r$
between local field and radial direction. Finally, we compute
the PDFs of these new angles and show them in dotted lines 
in Figure 12. The modeled PDFs of $\a_r$ appear to fit the 
measured PDFs reasonably well as soon as $\xi > 26\degr$. 
At the highest turbulence levels, above $\xi \approx 60\degr$, 
we note that the $\a_r$ PDFs, both measured and modeled, 
are nearly flat on a broad range of angles $\a_r$ between 
$50-60\degr$ and $110-120\degr$. This we now believe can be 
attributed to the broad and nearly axisymmetric distribution 
of the turbulent fields about the background-field direction.

\begin{figure} 
\epsscale{1.15} \plotone{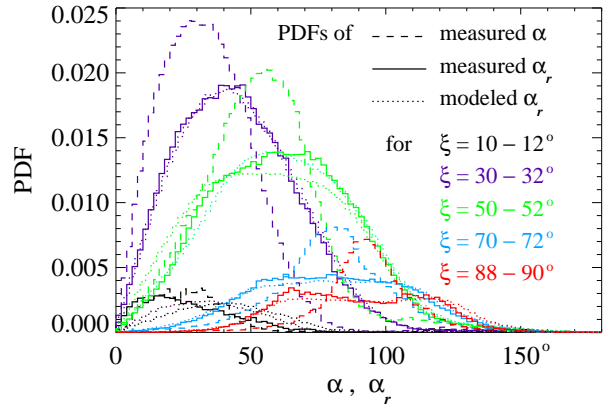} \vspace*{-.2cm} 
\caption{PDFs of measured $\a$ angles between local and background fields 
(dashed lines) and of measured and modeled $\a_r$ angles between local
field and radial direction (solid and dotted lines, respectively) for 
a series of turbulence levels $\xi = 11 \pm 1$\degr, 
$\xi = 31 \pm 1$\degr, $\xi = 51 \pm 1$\degr, $\xi = 71 \pm 1$\degr\ and
$\xi = 89 \pm 1$\degr. The green PDFs for $\xi = 51 \pm 1$\degr\ are
normalized and serve as reference level for the other ``PDFs.'' The model 
PDFs are obtained from the measured $\a$ PDFs by Monte-Carlo simulations 
assuming axisymmetry of the local field directions relative to the 
background. The dotted lines of matching color are obtained for a 
background field along the Parker spiral. The two additional dotted 
lines (green-blue for $\xi = 51\pm 1$\degr\ and dark-violet for 
$\xi = 11 \pm 1$\degr) are computed for a background field 
at 7\degr\ and 8\degr\ to the Parker spiral, respectively, 
slightly improving the fit to the measured PDFs. \\[-.0cm] } 
\label{fig12} 
\end{figure} 

In the cases $\xi = 50 \pm 1\degr$ and $\xi = 10 \pm 1\degr$, 
where the modeled PDFs are less than satisfactory, we do 
an additional transformation with a slightly different 
background field direction, shifted by some 7 and 8\degr\ 
relative to the Parker spiral. 
The results are also shown in dotted lines, but
with a slightly different color. They appear to improve
the fit to the measured PDFs of $\a_r$, but the fit
at the lowest turbulence level remains less than satisfactory.
The issue there again might be connected to the measured
``compressibility'' of the turbulence. 

We also note that the statistics in the central power bin 
($\xi = 51 \pm 1\degr$) are relatively high, 
and that a large number of SW streams contribute to these
statistics, with a spread in the Parker spiral direction of more 
than 4 degrees around the mean value of $\phi_r \approx 34\degr$. 
It may be affecting our result, but reduced statistics with more 
constrained values of $\Vsw$ and therefore $\phi_r$ did not seem 
to change that result. This issue will need further investigation. 

From the simple test of this section, we conclude that axisymmetry 
of the turbulent fields around a background field in or near the 
direction of the Parker spiral is a reasonably good assumption that 
produces fairly accurate fits of the measured PDFs of $\a_r$ angles, 
away from the lowest turbulence levels where $B_z$ turbulence becomes
dominant, that is, above $\xi \approx 26\degr$. This ``axisymmetry'' 
result appears to us difficult to reconcile with an $\a_r$ anisotropy 
of turbulence power driven by a wavevector anisotropy. 

\newpage

\section{CONSEQUENCES OF THE TURBULENCE-LEVEL \\[.1cm] 
VARIABILITY ON FIELD ORIENTATIONS \\[.1cm] AND APPARENT ANISOTROPY}

Figure 13 displays the PDFs of ``turbulence levels'' $\xi$ for 
series of angles $\a$ (left panels) and $\a_r$ (right panels)
measured between 0 and 90\degr\ (top panels) and between 0 and 180\degr\ 
(bottom panels). The left panels show PDFs peaking at the highest levels 
of turbulence near the normal to the background field (green PDF), 
and at the lowest levels near the direction of the background field 
(black, yellow), with a clear shifting of the peak with the angles 
$\a$ and $\a_{90}$. Magnetic fields near the normal to the background 
dominate at the highest levels of turbulence, while magnetic fields close 
to the background field direction dominate at the lowest turbulence levels 
$\xi$, which can also be seen in Figure 12 (see also dashed-line 
histograms for entire data set of Figure 9). As a consequence, because 
the angle between background and radial direction is relatively small 
in fast SW ($\sim 30\degr$), it is also true that magnetic fields at 
large angles to the radial are preponderant at higher $\xi$, while 
magnetic fields closer to the radial are more frequent at lower $\xi$. 
But the effect is less striking in the right panels of Figure 13 than 
that for the angles $\a$ relative to the background in the left panels. 

\begin{figure}[b] 
\epsscale{1.} \plotone{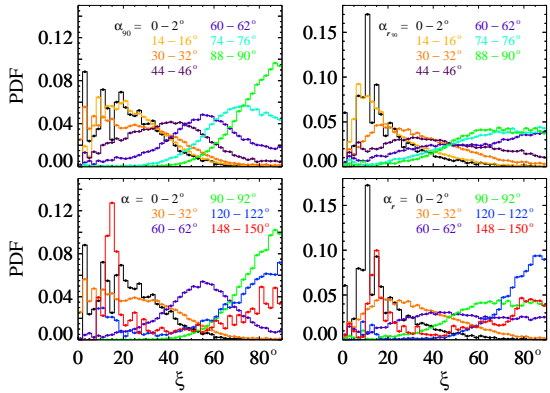} 
\caption{Measured PDFs of turbulence levels $\xi$ for series of 
intervals of angles between local and background fields, $\a_{90}$ and 
$\a$ (left panels), and between local field and radial, $\a_{r_{90}}$
and $\a_r$ (right panels). The angles $\a$ and $\a_r$ are measured 
between 0 and 180\degr, while the angles with an index 90 are measured 
between 0 and 90\degr. The left panels show PDFs peaking at the 
highest levels of turbulence near the normal to the background field
(green PDF), and at the lowest levels near the direction of the background 
field (black, yellow), with a clear shifting of the peak with the angles 
$\a$ and $\a_{90}$.} 
\label{fig13} 
\end{figure} 

\begin{figure}
\epsscale{1.} \plotone{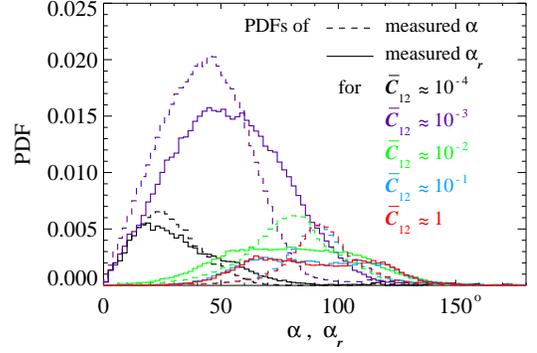} \vspace*{-.3cm}
\caption{``Same as'' Figure 12, but for a turbulence level defined as
$C_{12}/(\Bzav^2 \D z)$ (i.e., from the fluctuations at a single frequency
$\nu \approx 0.028\,$Hz) instead of 
$\langle (\D r)^2 \rangle^{}_{\Bzav} / (\D z)^2$, which depends on
a broad range of frequencies. The parameter $\overline{C}_{12}$ is 
$C_{12} / (2 \Bzav^2 \D z)$ and the PDFs are computed for the 
values of $\overline{C}_{12}$ giving, for $n=4$, 3, 2, 1 and 0, 
a $\xi_{12} \equiv \arctan[20 (2 \overline{C}_{12})^{1/2}]$ within 
a $2\degr$ bin containing $\arctan(20 \times 2^{1/2} \times 10^{-n/2})$
(the bins all start at multiples of $2\degr$). 
The violet PDFs for $\overline{C}_{12} = 10^{-3}$ are normalized 
and serve as reference level for the other ``PDFs.''}
\label{fig14} 
\end{figure} 

The strong ``anisotropy'' shown in Figures 12 and 13, and further in 
Figure 17, of course is related to our choice of the parameter $\xi$ 
to characterize the turbulence level [see Equations (\ref{C}--\ref{xi})]. 
This parameter closely orders the peaks of the $\a$ and $\a_{90}$ PDFs 
because it is related to the mean cross-field displacement of the 
turbulent field lines. But we argue that the power level has to be 
divided by the squared magnitude of the background magnetic field, 
as it is done in the definition of $C$ and therefore $\xi$, because 
in itself, the power in $\d B$ tells us nothing about turbulence; 
it is the power in $\d b \equiv \d B / \Bzav$ (of the fluctuations
relative to the background) that does.
No matter how high the power in $\d B$, if $\Bzav$ is high
enough, it will ``quench'' the turbulence. 

\begin{figure}
\epsscale{1.} \plotone{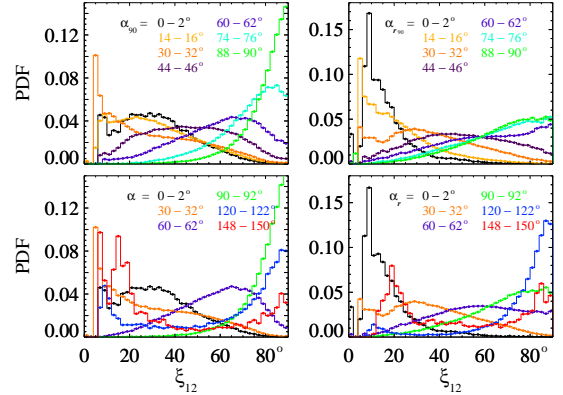} 
\caption{``Same as'' Figure 13, but for a turbulence level defined as
$C_{12}/(\Bzav^2 \D z)$ instead of $\langle (\D r)^2 \rangle_{\Bzav}/(\D z)^2$.
The parameter $\xi_{12}$ is $\arctan[20 \, (C_{12}/\D z)^{1/2}/\Bzav]$.} 
\label{fig15} 
\end{figure} 

\begin{figure}[b]
\epsscale{.8} \vspace*{-.1cm}\plotone{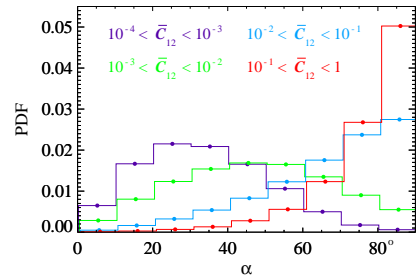} \vspace*{-.3cm}
\caption{Same as dotted-line histograms of the top panel of Figure 9, 
but for a turbulence level defined as $C_{12}/(\Bzav^2 \D z)$. The \\[-.1cm]
parameter $\overline{C}_{12}$ is one half that turbulence level. Again, 
the highest turbulence levels are found near the normal to the Parker 
spiral and the lowest levels near the Parker spiral direction.} 
\label{fig16} 
\end{figure} 

Figures 14 and 15 are the ``equivalent'' of Figures 12 and 13 
with a turbulence level defined using the field fluctuations $C_{12}/\Bzav^2$
at only one frequency (close to $\Vsw/\D z$), rather than 
using the mean cross-field displacement $\langle (\D r)^2\rangle^{1/2}_{\Bzav}$
as in Section 3.5. Similar conclusions can be drawn from these figures.
So our conclusion that the highest turbulence levels are found near 
the normal to the Parker spiral and the lowest levels near the 
Parker spiral direction is {\em not} due to the use of 
$\langle (\D r)^2\rangle^{}_{\Bzav}/(\D z)$ in place of $C_{12}/\Bzav^2$
in the definition of the turbulence level. Figure 16 also shows 
the PDFs of the angles $\a$ for a series of intervals of the 
parameter $\overline{C}_{12} \equiv C_{12}/(2 \Bzav^2 \D z)$. The PDFs 
of Figure 16 are to be compared to the dashed-line histograms of the 
top panel of Figure 9 for the entire data set. Again, the results 
are similar whether the definition of the turbulence level involves 
$\langle (\D r)^2\rangle^{}_{\Bzav}/(\D z)$ or only $C_{12}/\Bzav^2$. 

We note, however, that the parameter 
$\xi \equiv \arctan$ $(\langle (\D r)^2\rangle^{1/2}_{\Bzav} / \D z)$ 
used throughout this paper is a more meaningful ordering parameter
because the turbulent magnetic field lines are affected by the 
turbulent field fluctuations at a broad range of frequencies,
down to where the spectrum becomes flatter than $k^{-1}$, and 
not just by the fluctuations at the frequency inverse of the 
averaging timescale for the computation of the mean local 
field direction. The parameter $\xi$, in all instances where 
the transverse fluctuations dominate, that is, above about $26\degr$,
gives the approximate location of the peak in the PDFs of the angles 
$\a$ between mean local and background Parker fields.\footnote{Note 
again that if the turbulence were self-similar with only one 
``constant-at-all-scales'' spectrum of magnetic fluctuations, 
then $\langle (\D r)^2 \rangle$ and $C$ would simply be 
proportional to any one of the amplitudes $C_l$, that is, 
to what is ordinarily understood as the energy level of 
turbulence in self-similar, non-intermittent turbulence. 
It is only because the SW turbulence is strongly intermittent 
that involving the mean cross-field displacement 
$\langle (\D r)^2 \rangle^{1/2}$ in the definition of
the turbulence level presents an advantage.}

\begin{figure} 
\epsscale{1.04} \plotone{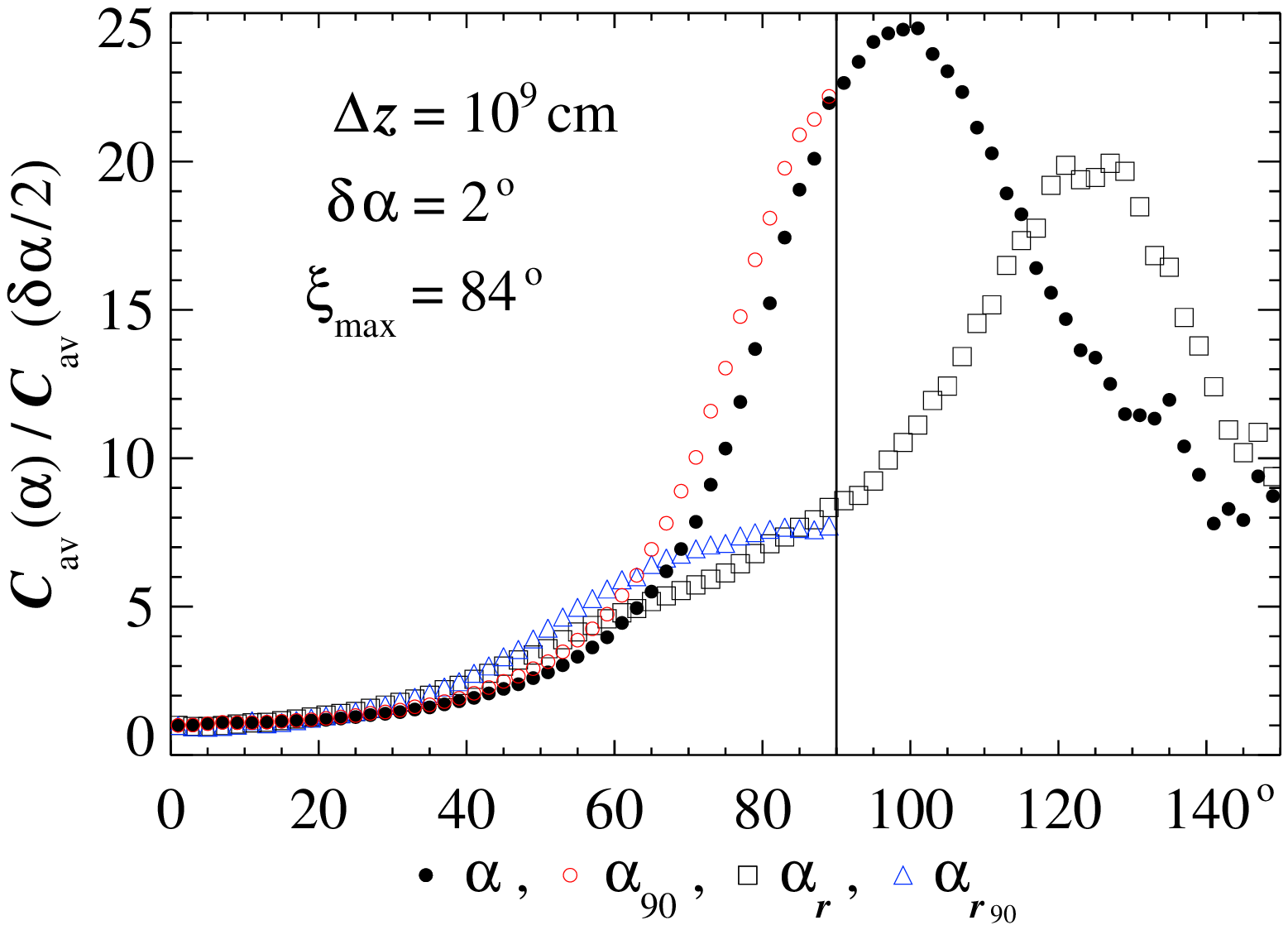} \vspace*{-.3cm}
\caption{Average turbulence levels $\Cav \equiv 
\int_{0^o}^{84^o} \!\! d\xi \, P(\xi, \a) \tan^2\xi$,  \\[-.05cm]
where $\tan \xi \equiv \langle (\D r)^2 \rangle^{1/2}_{\Bzav} / \D z$, 
relative to their value at the lowest \\[-.05cm]
angle $\d \a / 2 = 1$\degr, as functions of the angle $\a$ 
$\,$(filled$\,$ black$\,$ circles), \\[.05cm]
$\a_{ 90}\!$ (empty red circles), $\a_r\!$ (black squares)
and $\a_{r_{ 90}}\!\!$ (blue triangles).}
\label{fig17} 
\end{figure} 

\vspace*{.5cm}
Knowing the bivariate PDFs $P(\xi, \a')$ for $\a' = \a$, $\a_{90}$,
$\a_r$ and $\a_{r_{90}}$, we can now estimate the average turbulence
level at each angle $\a'$ by computing the integral
\eqb
\Cav(\a') \equiv \int_{0{\tiny \degr}}^{\xi_{\mbox{{\tiny max}}}} 
d\xi \, P(\xi, \a') \tan^2\xi \; , 
\label{Cav}
\eqe
with a cutoff at $\xi = \ximax$. The results relative to $\Cav(\d \a/2)$
in the first bin of the $\a'$ histograms, are presented in Figure 17
for $\ximax = 84\degr$. The peaks' heights are very sensitive to 
the value of the cutoff $\ximax$. For instance, a value 
$\ximax = 88\degr$ would produce a peak in $\a \approx 100\degr$
over 4 times higher than found here in Figure 17, and if all turbulence 
levels present in our data set were included, that peak would exceed 500. 
The value $\ximax = 84\degr$ was chosen because it is already sufficiently
large to produce high peaks, yet still low enough to keep 
the rising phase of the curves at low $\a'$ clear. 

Most importantly, we note that the ratio $\Cav(89\degr)/\Cav(1\degr)$ is 
much greater for $\a$ and $\a_{90}$ than it is for $\a_r$ and $\a_{r_{90}}$. 
The dependence on $\a$ of the turbulence level is the main effect,
that of the turbulence level on $\a_r$ a mere consequence of it. 
The steeper dependence on $\a$, or stronger ``anisotropy'' in $\a$, is due 
to the fact that the $\a$ PDFs remain peaked at all turbulence levels~$\xi$, 
whereas the $\a_r$ PDFs become flat at the higher $\xi$ (see Figure 12) 
as a result of the broad and presumably nearly axisymmetric distribution 
of field directions about the background field direction. 

\begin{figure}
\epsscale{1.04} \plotone{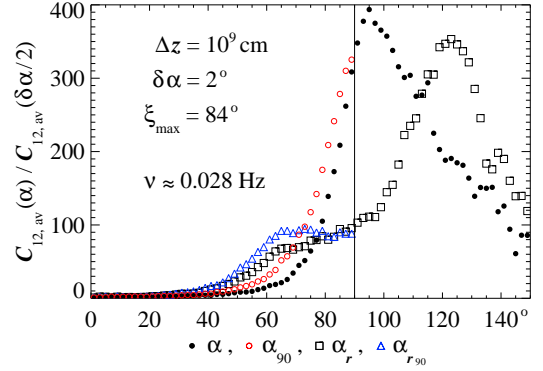} \vspace*{-.2cm}
\caption{Same as Figure 17, but with 
``$\tan \xi$'' defined as $(C_{12}/\D z)^{1/2} / |\Bzav|$ rather than 
$\langle (\D r)^2 \rangle^{1/2}_{\Bzav} / \D z$. The coefficient $C_{12}$
corresponds to a frequency $\nu \approx 0.028\,$Hz.}
\label{fig18} 
\end{figure} 

\begin{figure}
\epsscale{1.04} \plotone{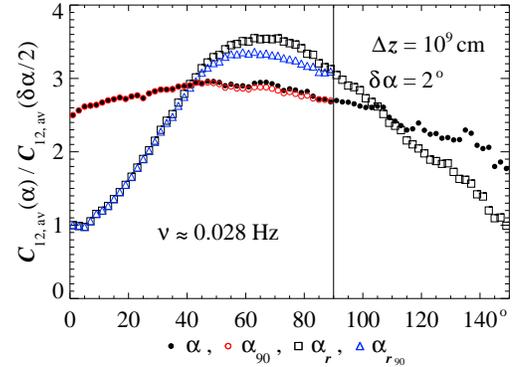} \vspace*{-.2cm}
\caption{Same as Figure 18, but with the long-time average $B_0$
substituting for the $15\,$min average $\Bzav$. Also, all 
$C_{12, \mbox{{\small av}}}$ are normalized to the same value of 
$C_{12, \mbox{{\small av}}}(\th_r)$ at $1\degr$ rather than their own 
value at $1\degr$. \\[-.2cm]}
\label{fig19} 
\end{figure} 

Also, the use of the amplitudes $C_l/(\Bzav^2 \D z)$ rather than $C$ 
as defined in Equation (\ref{C}) would lead to much steeper curves
and higher peaks, depending on the frequency (see Figure 18). 
The higher the frequency, the higher the peaks, 
consistent with the findings of Podesta (2009). 
Finally, when the running $15\,$min average $\Bzav$ of the background 
field magnitude is replaced by the long-time average $B_0$ in the
definition of $\xi$, the sharp dependence of the average turbulence
level $C_{12, \mbox{{\small av}}}$ on the angles $\a$ and $\a_r$ is 
damped out. It is reduced in Figure 19 to a mere factor $3-3.5$ 
increase in $\a_r$, and is almost entirely suppressed in $\a$. 
The result of Figure 19 for the dependence on $\a_r$ is very 
much consistent with the results shown in the top panel
of Figure 7 in Podesta (2009). 

The strong ``anisotropy'' presented in this section is in no way
related to any anisotropy in the wavevector distribution of the
turbulence. The fact that the turbulence levels are observed 
to be much higher in the directions close to the normal to 
the Parker spiral than in the directions close to the Parker
spiral itself is strictly due to the facts that (1) the field directions
deviate more strongly from the Parker spiral direction when the 
turbulence is higher, and (2) the turbulence level varies strongly
with time, enabling the observation of a wide range of field 
directions and turbulence levels. So the strong apparent 
``anisotropy'' presented in this section is strictly 
a consequence of the turbulence-level time variability.

\vspace*{.1cm}


\section{CONCLUSION}

\vspace*{.1cm}
In an effort to identify the effects of the broad variations in 
turbulence levels on the orientations of the local mean fields, we have 
analyzed the turbulent magnetic fields of a large set of fast SW 
streams measured onboard {\em ACE} and {\em STEREO A} and {\em B}. 
Our multiscale Fourier analysis of the transverse turbulent fields 
reveals time variations of the spectral power at all frequencies 
and on many timescales. The higher the time resolution of the 
spectral analysis is, the stronger the fluctuations of the 
power level are. The higher-resolution fluctuations far exceed 
the lower-resolution fluctuations, but roughly follow them 
in their time average (see Figures 1$-$3). 
From this multiscale Fourier analysis, we estimate as a linear 
combination of the Fourier amplitudes $C_l$ (see Figure 3) the 
GQL mean cross-field displacement of the magnetic field lines 
in the time-varying intermittent turbulence. This estimate of
the mean cross-field displacement in intermittent turbulence
is given by the square root of Equation (\ref{Dr}) with a local 
medium-scale ($\sim 15\,$min) average background field $\Bzav$ substituting 
for the stream average $B_0$. It includes the effects of all turbulent scales 
and fluctuates on many timescales. This estimate of the mean cross-field 
displacement, once divided by the field-aligned scale $\D z$, 
defines the square root $C^{1/2}$ of the power level of the 
intermittent turbulence in Equation (\ref{C}), and the related 
angle, $\xi \equiv \arctan C^{1/2}$, used throughout the paper 
to parametrize the power level of the intermittent turbulence. 

Provided ``compressibility effects'' are included at the lowest
power levels of turbulence ($\xi < 26\degr$), modeling the PDFs of 
the angles $\a$ between local and background fields produces 
satisfactory fits of the observed PDFs at all stable power levels 
of turbulence (see Figures 8, 10 and 11), that is, on all time 
intervals with variations in the power level of turbulence that 
are sufficiently slow. Because the highest power levels of 
turbulence happen in short bursts, the statistics for stable 
power levels are biased toward the lower turbulence levels 
and smaller angles (see Figures 6$-$9). But the fits remain 
quite accurate at all levels $\xi$, even for the very low 
statistics of the higher $\xi$. Because our GQL modeling 
does not assume anything about the three-dimensional distribution
of wavevectors in the turbulence, and therefore does not preclude 
an isotropic distribution of wavevectors, it follows from these 
fits that an isotropic turbulence could account for the measured 
PDFs of the angles $\a$ between local and background fields
at all stable levels of turbulence.

Both the direct measure of the field projection on the 
background field (Parker spiral direction) and the PDFs of 
the measured angles $\a$ between local and background fields
reveal local field reversals that are quite common even
within very broad streams of ``unipolar'' fast SW (see Figures 5 and 12). 
On the whole, they happen some 4 to 10 percent of the time, 
but at the highest levels of turbulence, short-scale reversals
are so frequent that on the scale of $10^9\,$cm, reversed 
fields are practically as common as normal-polarity fields.

Modeling through Monte-Carlo simulation the PDFs of the angles 
$\a_r$ between local fields and radial direction from the 
measured PDFs of the angles $\a$, we find that axisymmetry of 
the turbulent fields around a background field in or near the 
direction of the Parker spiral is a reasonably good assumption. 
The modeling made under this simple assumption does indeed 
produce fairly accurate fits of the measured PDFs of $\a_r$ 
angles, away from the lowest turbulence levels where $B_z$ 
fluctuations become dominant, that is, above $\xi \approx 26\degr$.
At the highest turbulence levels, above $\xi = 60\degr$, both the 
model and observed $\a_r$ PDFs are nearly flat on a broad range 
of angles $\a_r$ between $50-60\degr$ and $110-120\degr$ (see Figure~12). 

Unsurprisingly, magnetic fields near the normal to the background 
field or Parker spiral dominate at the highest turbulence levels,
while magnetic fields close to the Parker spiral direction 
dominate at the lowest turbulence levels $\xi$, 
with a peak of the $\a$ ($\xi$) angle PDFs smoothly shifting from 
the parallel to the normal direction (from the low to the high 
turbulence levels) as $\xi$ ($\a$) increases [Figure 12 (13)]. 
This results in a very steep dependence of the average power level 
of turbulence $\Cav$ of Equation (\ref{Cav}) on the angle to the 
Parker spiral, and in a more moderate dependence on the angle to 
the radial (Figure~17). The steeper dependence on $\a$ is due to 
the fact that the $\a$ PDFs remain peaked at all $\xi$ whereas
the $\a_r$ PDFs, due to the broad and nearly axisymmetric PDFs of the
field directions relative to the background, become flat at the 
higher $\xi$. The dependence is found to be even steeper for the 
higher-frequency amplitudes $C_{12, \mbox{{\small av}}}/\Bzav^2$ (Figure 18), 
but strongly reduced for $C_{12, \mbox{{\small av}}}$ (Figure 19).
Our conclusion that  magnetic fields near the normal to the background 
Parker spiral dominate at the highest turbulence levels, while 
magnetic fields close to the Parker spiral direction dominate 
at the lowest turbulence levels is not modified by the use of 
a single-frequency turbulence-level definition $C_{12}/(\Bzav^2 \D z)$
in the place of $\langle (\D r)^2\rangle^{}_{\Bzav} / (\D z)^2$ (Figures~14--16).

We do not presume to know whether the $\a_r$ ``anisotropy'' 
of $C_{12, \mbox{{\small av}}}$ presented in Figure 19 is a 
residual effect of the strong variability in the power level 
of turbulence. But we can certainly conclude that the extreme
``anisotropy'' seen in Figures 17 and 18 is unambiguously caused
by the time variability in that power level of turbulence, and not 
by any anisotropy in the wavevector distribution of the turbulence. 
Clearly, a purely isotropic model of turbulence, that is, 
a model of turbulence with an isotropic distribution of 
wavevectors $\vk$, {\em can} reproduce the PDFs of the angles
$\a$ and $\a_r$ between local and background fields and 
between local field and radial direction presented in this
paper for series of the power level of turbulence. 
By this we mean that an isotropic model of turbulence cannot
be excluded on the basis of the observed dependences of the 
angle PDFs on the power level of turbulence, or of the average 
power levels of turbulence on the angles. These dependences 
are entirely imputable to the time variability in the power level
of turbulence observed in the fast SW streams. These results
are consistent with the near isotropy found by Narita et al. (2010) 
from four-point magnetic field measurements of the Cluster spacecraft 
at the turbulence scales close to $10^9\,$cm in the SW. 

The parameter $\xi$ defined by Equations (\ref{C}) and (\ref{xi}) does 
order the directions of the local fields best because it physically 
represents the local {\em statistical} average directions. This tight 
ordering of the local field directions with the parameter $\xi$ brings 
the observed ``anisotropies'' of the average power level of turbulence 
to extremes much higher than otherwise found with the power level
{\em undivided} by the local background field. But again, we argue 
that the power level has to be divided by the squared magnitude 
of the background magnetic field because in itself, 
the power in $\d B$ tells us nothing about turbulence, 
it is the power in $\d b \equiv \d B / \Bzav$ that does.
No matter how high the power in $\d B$, if $\Bzav$ is high
enough, it will ``quench'' the turbulence. 

The variations of $\Bzav$ itself are likely caused, in large part, 
by the non-uniformity of the emerging fields within each of the 
coronal holes at the source of the fast SW streams. We suggest that 
some of the variations of $\Bzav$, in particular the local field 
reversals (Figure 5; Section 3.6), may also be signs of intermixed 
magnetic fields of opposite polarity, originating from a different 
SW stream, perhaps even signs of ongoing reconnection with these 
opposite-polarity fields, or remnants of reconnections that took place 
earlier. Though it is of course also possible and even likely that some 
of these local field reversals are due to the reconnection, at the basis 
of the corona, within coronal holes, of open field lines 
with closed magnetic loops.

\acknowledgments 
 
This research was funded by NSF under the grant 0940976 of the 
Solar Terrestrial Research Program. The {\em ACE} MAG and 
{\em STEREO} IMPACT/MAG data \linebreak were downloaded from the 
{\em ACE} science center at \linebreak www.srl.caltech.edu/ACE/ASC 
and from sprg.ssl.ber\-keley.edu/impact.


\end{document}